\documentclass[reprint,prd,showpacs,preprintnumbers,amsmath,amssymb,superscriptaddress]{revtex4-1}
\usepackage{graphicx}
\usepackage{dcolumn}
\usepackage{amsmath}
\usepackage{times}
\usepackage{mathtools}
\usepackage{color}

\begin{document}

\title{Treatment of broken symmetry in the Faddeev approach to the strange baryon spectra  }

\author{Joseph P.\ Day}
\affiliation{ Institute for Theoretical Physics, University of Graz,  A-8010 Graz, Austria}

\author{Zolt\'an  Papp}
\affiliation{ Department of Physics and Astronomy,
California State University Long Beach, Long Beach, California, USA }

\date{\today}

\begin{abstract}
By solving the Faddeev equations we calculate the mass of the strange baryons in the framework of a relativistic constituent quark model. The Goldstone-boson-exchange quark-quark interaction is derived from $SU(3)_F$ symmetry, which is explicitly broken as the strange quark is much heavier. This broken symmetry can nicely be accounted for in the Faddeev framework.

\end{abstract}


\maketitle

\section{Introduction}
\label{intro}
In the relativistic constituent quark model most of the strange baryons are made of light and strange quarks. In these models the quarks are considered to be identical particles. The interaction is derived from the exchange of particles, like the exchange of Goldstone bosons. {However, due to the explicit breaking of chiral symmetry the quarks acquire mass, which can be quite different}. 

In this work we present results from a study of strange baryon spectra within a relativistic framework. In particular, we demonstrate the performance of the Faddeev approach  {with respect to a relativistic constituent quark model (RCQM), where the constituent quarks acquire a dynamical mass due to their collective interactions with the QCD vacuum. In $SU(2)_F$ the symmetry is almost exact but  with the incorporation of the strange quark, due to explicit symmetry breaking, $SU(3)_F$ becomes only approximate}. The up and down quarks are treated as equal mass particles within the framework of isospin formalism while the strange quark is treated, as seen in nature, as a heavier particle. The relativistically invariant mass spectra is obtained by a Faddeev integral-equation method, adapted to treating long-range interactions, such as the quark confinement. In the Faddeev approach the wave function is broken into three components.  If some pair of particles are symmetric, then the corresponding Faddeev components have the same functional form. This allows us to reduce the number of components. Here we treat the two light and one strange  or a two strange and one light quark system in a two-component Faddeev model. One component is responsible for the light-light or strange-strange pair, while the other one for the light-strange pair.  {This way we retain the $SU(3)_F$ limit as the interactions are derived from this symmetry, but due to the explicit symmetry breaking of the system, the masses of the constituent quarks are different,  hence we do not impose any explicit symmetry between strange and light quarks}.

In Section \ref{sec2} below we outline the Faddeev approach to three-quark systems with a confining long range potential.
In Section \ref{sec3}  {we discuss the consequences of the permutation symmetry of the particles on the formalism}. Then in Section \ref{sec-gbe} we sketch the Goldstone-boson-exchange model. 
{This model is inspired by the work of Glozman and Riska \cite{Glozman:1995fu} and it has subsequently 
been elaborated thoroughly by Glozman, Plessas and collaborators of the Graz group 
(see eg.\ Refs.\ \cite{Glozman:1996wq,Glozman:1997fs,Glozman:1997ag,melde08} and references therein).}
In Section 
\ref{gbe-res} we present the results of our calculations and finally we draw some conclusions.

\section{Faddeev approach to three-quark problems }
\label{sec2}

We consider a three-particle Hamiltonian 
\begin{equation}\label{hamilton}
H=H^{(0)}+v_{1}+v_{2}+v_{3},
\end{equation}
where $H^{(0)}$ is the kinetic-energy operator and $v_{\alpha}$,  with $\alpha=1,2,3$, are the 
mutual interactions of the quarks. We represent it through the usual configuration-space
Jacobi coordinates: e.g.\ ${\vec x}_{1}$ is the coordinate between particles $2$ and $3$ and ${\vec y}_{1}$ is 
the coordinate between the center of mass of the pair $(2,3)$ and particle $1$. 

The kinetic energy operator is given in the relativistic form
\begin{equation}
H^{(0)}=\sum_{i=1}^{3}\sqrt{k_{i}^{2}+m_{i}^{2}},
\label{relh0}
\end{equation}
where $m_{i}$ are the quark masses and ${\bf k}_{i}$ are the three-momenta of the quarks in the reference frame where the total three-momentum
${\bf P} = \sum_{i=1}^{3} {\bf k}_{i} =0$. 

The quark-quark interaction is a long range confining potential.  In order that we can apply the Faddeev procedure along the method presented in 
Refs.\  \cite{Papp:1998yt,Papp:2000kp,McEwen:2010sv} we split the quark-quark potential
into confining and non-confining parts
\begin{equation}
v_{\alpha}= v_{\alpha}^{(c)} + v_{\alpha}^{(s)},
\end{equation}
where superscripts $c$ and $s$ stand for confining and short-range, respectively.
Then the Schr\"odinger equation takes the form
\begin{equation}\label{schrodinger}
H|\Psi\rangle=(H^{(c)}+v_{1}^{(s)}+v_{2}^{(s)}+v_{3}^{(s)} )|\Psi \rangle = E |\Psi\rangle~,
\end{equation} 
with 
\begin{equation}
H^{(c)}= H^{(0)}+v_{1}^{(c)}+v_{2}^{(c)}+v_{3}^{(c)}.
\end{equation}
We introduce
\begin{equation}
G^{(c)}(E)=(E-H^{(c)})^{-1},
\end{equation}
 and by rearranging (\ref{schrodinger}),  we have
\begin{eqnarray}\label{fdec1}
|\Psi\rangle && =  G^{(c)}(E)(v_{1}^{(s)}+v_{2}^{(s)}+v_{3}^{(s)})|\Psi \rangle \\
&& = G^{(c)}(E)v_{1}^{(s)}|\Psi \rangle + 
G^{(c)}(E) v_{2}^{(s)} |\Psi \rangle +G^{(c)}(E)v_{3}^{(s)}|\Psi \rangle. \nonumber
\end{eqnarray} 
So,  the three-particle wave function $|\Psi\rangle$ naturally splits into three 
components 
\begin{equation}\label{fdec2}
|\Psi\rangle=|\psi_{1} \rangle + |\psi_{2} \rangle + |\psi_{3} \rangle~,
\end{equation} 
where
\begin{equation}\label{fcomp}
|\psi_{\alpha}\rangle=G^{(c)}(E)v_{\alpha}^{(s)}|\Psi \rangle, \ \ \ \ \alpha=1,2,3,
\end{equation}
are the Faddeev components. They satisfy the set of equations, the Faddeev equations,
\begin{eqnarray}\label{faddeqdiff}
(E-H^{(c)}-v_{1}^{(s)})|\psi_{1} \rangle &=& v_{1}^{(s)}( |\psi_{2} \rangle + |\psi_{3} \rangle) \nonumber \\
(E-H^{(c)}-v_{2}^{(s)})|\psi_{2} \rangle &=& v_{2}^{(s)}( |\psi_{1} \rangle + |\psi_{3} \rangle) \nonumber \\
(E-H^{(c)}-v_{3}^{(s)})|\psi_{3} \rangle &=& v_{3}^{(s)}( |\psi_{1} \rangle + |\psi_{2} \rangle)~. 
\end{eqnarray} 
Indeed, by adding up these equations and considering (\ref{fdec1}) we recover the Schr\"odinger
equation. 
With the help of channel Green's operators 
\begin{equation}
G_{\alpha}^{(c)}(E)=(E-H^{(c)}-v_{\alpha}^{(s)})^{-1},
\end{equation}
 we can rewrite Eqs.\ (\ref{faddeqdiff}) into an integral equation form,
\begin{eqnarray}\label{faddeqint}
|\psi_{1} \rangle &=& G_{1}^{(c)}(E) \: v_{1}^{(s)}( |\psi_{2} \rangle + |\psi_{3} \rangle) \nonumber \\
|\psi_{2} \rangle &=& G_{2}^{(c)}(E)\: v_{2}^{(s)}( |\psi_{1} \rangle + |\psi_{3} \rangle) \nonumber \\
|\psi_{3} \rangle &=& G_{3}^{(c)}(E)\: v_{3}^{(s)}( |\psi_{1} \rangle + |\psi_{2} \rangle)~. 
\end{eqnarray} 
  
Even for such a simple system like a three-quark system the wave function $\Psi$ can be rather complicated. 
The quark-quark potential for various angular momentum channels may have strong attractive and repulsive components. This
leads to a strong quark-quark correlation in each of the possible subsystems of the three-quark system, which is hard to describe
by a single wave function.
The Faddeev components possess a simpler structure. For example, in Eq.\ (\ref{fcomp}), the short range potential $v_{1}^{(s)}$ acting on 
$|\Psi\rangle$, suppresses those asymptotic structures when particles $2$ and $3$ are far away.
Consequently,  $|\psi_{1}\rangle$ contains only one kind of physical situation when particle $1$ is far away and particles 
$2$ and $3$ are in strong correlation. A similar statement is valid for 
$|\psi_{2}\rangle$ and $|\psi_{3}\rangle$. So, with the Faddeev decomposition we achieve a splitting of the wave function into parts 
such that each component possesses only one kind of asymptotic behavior and represents one kind of two-quark correlations. 

We need to introduce the appropriate angular momentum basis. The angular momentum associated with coordinates $x_{\alpha}$ and 
$y_{\alpha}$ are denoted by $l_{\alpha}$ and $\lambda_{\alpha}$, respectively, and they are coupled to the total angular momentum $L$. The spin of
particles $\beta$ and $\gamma$, $S_{\beta}$ and $S_{\gamma}$, respectively, are coupled to $s_{\alpha}$, which is with 
the spin of particle $\alpha$, $S_{\alpha}$, 
coupled to the total spin $S$. Similarly, the isospin of
particles $\beta$ and $\gamma$, $t_{\beta}$ and $t_{\gamma}$, respectively, are coupled to $\tau_{\alpha}$, which is with the isospin of particle 
$\alpha$, $t_{\alpha}$, 
coupled to the total isospin $T$. The angular momentum $L$ and spin $S$ are coupled to total angular momentum $J$. So, 
we adopted $LS$ coupling, which is appropriate if the quark-quark interaction does not have tensor terms.

\section{Faddeev equations with identical particles}
\label{sec3}

A further advantage of the Faddeev method is that the identity of particles greatly simplifies the equations 
(see eg.\ Ref.\ \cite{Schellingerhout:1995gn}).
If particles $\beta$ and $\gamma$ are identical the wave function must be symmetric with respect to exchange of these particles. 
We denote ${\cal P}_{\alpha}$ the operator which exchanges particles $\beta$ and $\gamma$. Then 
\begin{equation}
{\cal P}_{\alpha}|\Psi \rangle = p_{\alpha}|\Psi\rangle
\end{equation}
where 
\begin{equation}
p_{\alpha}=(-1)^{l_{\alpha}+s_{\alpha}-S_{\beta}-S_{\gamma} + \tau_{\alpha} - t_{\beta} -t_{\gamma}}
\label{sym}
\end{equation}
if particles carry spin and isospin.
The Faddeev component for $\beta$ is defined by 
\begin{equation}
|\psi_{\beta}\rangle = G^{(c)} v_{\beta}^{(s)} |\Psi \rangle.
\end{equation}
By applying ${\cal P}_{\alpha}$ we have
\begin{equation}
{\cal P}_{\alpha}|\psi_{\beta}\rangle = {\cal P}_{\alpha} G^{(c)} v_{\beta}^{(s)} |\Psi \rangle =  
G^{(c)} {\cal P}_{\alpha} v_{\beta}^{(s)} {\cal P}_{\alpha}p_{\alpha} |\Psi \rangle.
\label{p23psi}
\end{equation}
Considering that $v^{(s)}_{\beta}$ is the interaction between particles $\alpha$ and $\gamma$ and the 
operator $P_{\alpha}$  exchanges particles $\beta$ and $\gamma$, we get
\begin{equation}
{\cal P}_{\alpha} v^{(s)}_{\beta} {\cal P}_{\alpha} = v^{(s)}_{\gamma},
\end{equation}
and Eq.\ (\ref{p23psi}) becomes
\begin{equation}
p_{\alpha}{\cal P}_{\alpha}|\psi_{\beta}\rangle = G^{(c)} v_{\gamma}^{(s)} |\Psi \rangle.
\end{equation}
The right hand side of this equation is the defining relation for $|\psi_{\gamma}\rangle$. So, we can conclude that
\begin{equation}
|\psi_{\gamma}\rangle = p_{\alpha} {\cal P}_{\alpha}|\psi_{\beta}\rangle,
\end{equation}
and in a similar manner
\begin{equation}
|\psi_{\beta}\rangle = p_{\alpha} {\cal P}_{\alpha}|\psi_{\gamma}\rangle.
\end{equation}
Since ${\cal P}_{\alpha}v_{\alpha}^{(s)} {\cal P}_{\alpha}=v_{\alpha}^{(s)}$, we get
\begin{equation}
{\cal P}_{\alpha}|\psi_{\alpha}\rangle = G^{(c)} {\cal P}_{\alpha}v_{\alpha}^{(s)} {\cal P}_{\alpha} p_{\alpha} |\Psi \rangle = p_{\alpha}  |\psi_{\alpha} \rangle.
\end{equation}
So, if particles $\beta$ and $\gamma$ are identical, then the angular channels for $|\psi_{\alpha}\rangle$ have to be selected such that
$p_{\alpha}=1$  for bosons and  $p_{\alpha}=-1$ for fermions. On the other hand, the Faddeev equation for $|\psi_{\alpha}\rangle$ is given by
\begin{equation}
|\psi_{\alpha}\rangle= G_{\alpha}^{(c)}v_{\alpha}^{(s)}(|\psi_{\beta}\rangle + |\psi_{\gamma}\rangle)=
 (1+p_{\alpha} {\cal P}_{\alpha})G_{\alpha}^{(c)}v_{\alpha}^{(s)}|\psi_{\beta}\rangle.
\end{equation}
We should notice that $1+p_{\alpha} {\cal P}_{\alpha}$ is just twice the symmetrizing or anti-symmetrizing operator. 
If we select the angular basis for $|\psi_{\alpha}\rangle$ such that it ensures
the correct symmetry or anti-symmetry the value of $1+p_{\alpha} {\cal P}_{\alpha}$ is just $2$. 

Putting everything together and assuming that particles $2$ and $3$  
are identical  the three-component Faddeev equation simplifies to
\begin{equation}
\begin{pmatrix}
|\psi_{1} \rangle \\
|\psi_{2} \rangle
\end{pmatrix} =
\begin{pmatrix}
0 & 2  G_{1}^{(c)} v_{1}^{(s)}  \\
G_{2}^{(c)}  v_{2}^{(s)} &  G_{2}^{(c)} v_{2}^{(s)} p_{1} {\cal P}_{1}
\end{pmatrix}
\begin{pmatrix}
|\psi_{1} \rangle \\
|\psi_{2} \rangle
\end{pmatrix}.
\label{2identicalfe}
\end{equation}

If all three particles are identical Eq.\ (\ref{2identicalfe}) gets further reduced  to one single equation
\begin{equation} \label{fmp}
| \psi_{1} \rangle =  2 G_1^{(c)} v_1^{(s)} {\mathcal P}_{123}
| \psi_{1} \rangle,
\end{equation}
where ${\mathcal P}_{123}={\mathcal P}_{12}{\mathcal P}_{23}$ is the operator for cyclic permutation of all three particles 
${\mathcal P}_{123} |\psi_{1}\rangle =| \psi_{2}\rangle$. 

The solution of the Faddeev equations for confining potentials has been presented in Refs.\ \cite{Papp:2000kp,McEwen:2010sv}. We found that the simpler and faster method of Ref.\ \cite{Papp:2000kp} and the more elaborated method
of Ref.\ \cite{McEwen:2010sv} provides results which are in a very good agreements with each other. Therefore, in this paper we adopted the simpler method. These methods entails a separable approximation of the short-range parts of the quark-quark potentials on the three-body 
basis. Our basis is defined as before 
\begin{equation}
\langle x_{\alpha} y_{\alpha}| n \nu \rangle_{\alpha} = \{ \langle x_{\alpha} | n \rangle \langle y_{\alpha} | \nu \rangle \},
\end{equation}
where $\langle x | n \rangle$ are the Coulomb-Sturmian functions. The bracket stands for angular momentum coupling and for simplicity 
we have suppressed the angular momentum, the spin and isospin indexes. In this method we need to evaluate numerically the matrix elements 
$_{1}\langle n \nu |v_{1}^{(s)}|n' \nu' \rangle_{2}$, $_{2}\langle n \nu |v_{2}^{(s)}|n' \nu' \rangle_{1}$ and
 $\mbox{}_{2}\langle n \nu |v_{2}^{(s)}|n' \nu' \rangle_{3}$, which can be done in configuration space representation. The Coulomb-Sturmian functions also have a nice
 analytic form in momentum space, which facilitates the evaluation of matrix elements of the Green's operator $G^{(c)}_{\alpha}$ as 
 presented  in Ref.\ \cite{Papp:2000kp}.

\section{Goldstone boson exchange model for baryons}
\label{sec-gbe}

We consider baryons with Goldstone boson exchange (GBE) quark-quark interaction
\begin{equation}
v_{\alpha}=V^{\text{conf}}_{\alpha}+V^{\chi}_{\alpha}~,
\end{equation}
where the confinement potential is taken in the form
\begin{equation}
V^{\text{conf}}_{\alpha}=V_{0}+C x_{\alpha}~.
\end{equation}
The chiral potential is a sum of octet
\begin{eqnarray}
V^{\chi}_{\alpha}(\vec{x}_{\alpha}) &=&
\sum_{F=1}^{3} V_{\pi} ({\vec x}_{\alpha})  \lambda_{\beta}^{F} \lambda_{\gamma}^{F} {\vec \sigma}_{\beta} {\vec \sigma}_{\gamma} 
\nonumber \\
&& + \sum_{F=4}^{7} V_{K} ({\vec x}_{\alpha})  \lambda_{\beta}^{F} \lambda_{\gamma}^{F} {\vec \sigma}_{\beta} {\vec \sigma}_{\gamma} 
\nonumber \\
&& + V_{\eta} ({\vec x}_{\alpha})  \lambda_{\beta}^{8} \lambda_{\gamma}^{8}   
 {\vec \sigma}_{\beta} {\vec \sigma}_{\gamma},
 \label{octet}
\end{eqnarray}
and singlet
\begin{equation}
V^{\chi}_{\alpha}(\vec{x}_{\alpha}) = 
 \frac{2}{3} V_{\eta'} ({\vec x}_{\alpha})   
 {\vec \sigma}_{\beta} {\vec \sigma}_{\gamma},
 \label{singlet}
\end{equation}
 meson-exchange terms, where ${\vec \sigma}$ and $\lambda^{F}$ are the quark spin and flavor matrices, respectively.
Terms $V_{\pi}$, $V_{K}$, $V_{\eta}$, and $V_{\eta'}$ represent the form factor
for $\pi$, $K$, $\eta$, and $\eta'$ meson exchanges, respectively. 
 {In this model we assume that in the $SU(3)_F$ chiral limit all quarks are equivalent, specifically up, down, and strange current quark masses can be identically set to zero. This symmetry is broken in the low energy limit of the theory where it is assumed that the baryon mass is only marginally dependent on the current quark mass due to the majority of the mass being determined by the spontaneous breaking of chiral symmetry and the effective confining interaction. The quarks acquire a constituent mass due to the spontaneous and dynamical breaking of chiral symmetry and, in addition, the light and strange quarks acquire different masses due to the explicit breaking of this symmetry.  However, we need to ensure the symmetry in the chiral limit. Therefore,
the exchange of Goldstone bosons should be symmetric}. This introduces a symmetry factor to the pion and kaon exchange terms.

So,  in angular momentum basis we get a flavor dependent quark-quark interaction for light quarks
\begin{equation}
V^{\text{u(d)-u(d)}}_{\alpha} = \left\{  p_{\alpha}^{u(d)-u(d)} V_{\pi} T_{\alpha}  +\frac{1}{3}V_{\eta}+ \frac{2}{3}V_{\eta'}\right\} 
\Sigma_{\alpha},
  \label{vuu}
\end{equation}
for the light and strange quarks
\begin{equation}
V^{\text{u(d)-s}}_{\alpha} =\left\{ 2 p_{\alpha}^{u(d)-s}  V_{K} -\frac{2}{3}V_{\eta}+ \frac{2}{3}V_{\eta'}  \right\} \Sigma_{\alpha}, 
\label{vus}
\end{equation}
and for the strange quarks
\begin{equation}
V^{\text{s-s}}_{\alpha} = \left\{ \frac{4}{3}V_{\eta}+ \frac{2}{3}V_{\eta'}  \right\} \Sigma_{\alpha}.
\label{vss}
\end{equation}
The quarks are spin $1/2$ particles and the isospin of the light quarks are $1/2$ while the isospin of the strange 
quarks equals to $0$. So for the symmetry coefficients we have
\begin{equation}
p_{\alpha}^{u(d)-u(d)}=(-1)^{l_{\alpha}+s_{\alpha}+\tau_{\alpha}-2}
\end{equation}
and
\begin{equation}
p_{\alpha}^{u(d)-s}=(-1)^{l_{\alpha}+s_{\alpha}-1},
\end{equation}
and the $T_{\alpha}$ and $\Sigma_{\alpha}$ isospin-isospin and spin-spin factors  are given by
\begin{equation}
T_{\alpha} =  2 \tau_{\alpha} (\tau_{\alpha}+1)-3
\end{equation} 
and 
\begin{equation}
\Sigma_{\alpha} =  2 s_{\alpha} (s_{\alpha}+1)-3,
\end{equation}
respectively.

In the GBE model the spatial parts of the potential is taken as sum of two Yukawa terms
\begin{equation}
V_{\gamma}= \frac{g^2_{\gamma}}{4 \pi} \frac{1}{12 m_i m_j} \bigg[ \mu^2_{\gamma} \frac{e^{-\mu_{\gamma} r_{ij}}}{r_{ij}} -\Lambda_{\gamma}^2 \frac{e^{-\Lambda_{\gamma}r_{ij}}}{r_{ij}}\bigg]
\end{equation}
for $\gamma =  \pi , K, \eta,$ and $\eta'$. The coupling constant $\Lambda_{\gamma}$ is assumed to have a linear dependency on meson masses
\begin{equation}
\Lambda_{\gamma}=\Lambda_{0}+\kappa \mu_{\gamma}~,
\end{equation}
where $\Lambda_{0}$ and $\kappa$ are free parameters. In this model the masses of quarks and the mesons are fixed parameters, $m_{u}=m_{d}=340$ MeV, 
$m_{s}=500$ MeV, $\mu_{\pi}=139$ MeV, $\mu_{K}=494$ MeV, $\mu_{\eta}=547$ MeV, $\mu_{\eta'}=958$ MeV, and the meson octet-quark coupling constant was 
adopted as $g_{8}^{2}/ 4\pi=0.67$. 
The other parameters for the model were determined by fitting manually to the observed baryon spectra. Excellent agreement with experiment was found with
$V_{0}=416$ MeV and $C=2.33\ {\mathrm fm}^{-2}$, $\Lambda_{0}=2.86\ {\mathrm fm}^{-1}$, $\kappa=0.81$ and $(g_{0}/g_{8})^{2}=1.34$ 
\cite{Glozman:1997ag}. 

It is natural to incorporate $V^{\text{conf}}$ into $v^{(c)}$ and $V^{\chi}$ into $v^{(s)}$. 
In Refs.\ \cite{Papp:1998yt,Papp:2000kp,McEwen:2010sv} we showed that in order to avoid the appearance of spurious solutions the splitting of the quark-quark potential to
$v^{(c)}$ and $v^{(s)}$ should be performed such that in the region of physical interest $G^{(c)}$ do not have poles.
To ensure this, we add a repulsive Gaussian term to $V^{\text{conf}}$, which we subtract from
$v^{\chi}$
\begin{equation}
v^{(c)}  =V^{\text{conf}} + a_0 e^{-(r/r_0)^2}
\end{equation}
and 
\begin{equation}
v^{(s)} = V^{\chi}  - a_0 e^{-(r/r_0)^2}\ .
\end{equation}
 The parameters of the auxiliary potential have been taken as 
$a_0=3\;\mbox{fm}^{-1}$ and $r_0=1\;\mbox{fm}$. By this choice of the
parameter
values any bound states of $H^{(c)}$ are avoided below $\approx 2$ GeV. The
values
of $a_0$ and $r_0$ also influence the rate of convergence, but not the
final  results. The other parameters of the calculations are the same as for the light baryons 
Ref.\ \cite{McEwen:2010sv}.

\section{The GBE results for strange baryons}
\label{gbe-res}

\begin{figure}[htb]
\includegraphics[width=8cm]{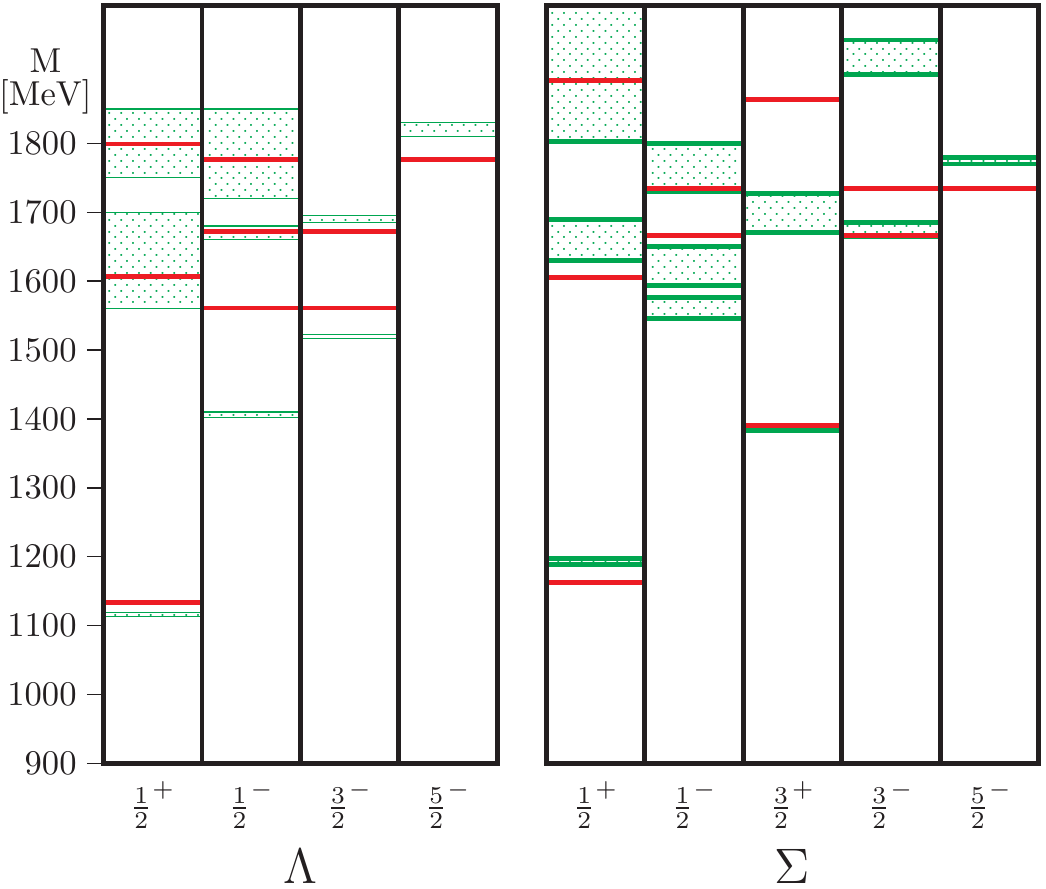}
\caption{\label{fig1} Faddeev calculations of $\Lambda$ and $\Sigma$ baryons are given by the red line. Experimental results as reported by the Particle Data Group \cite{pdg} with uncertainty are given by the green boxes.}
\end{figure}

\begin{figure}[htb]
\includegraphics[width=5.5cm]{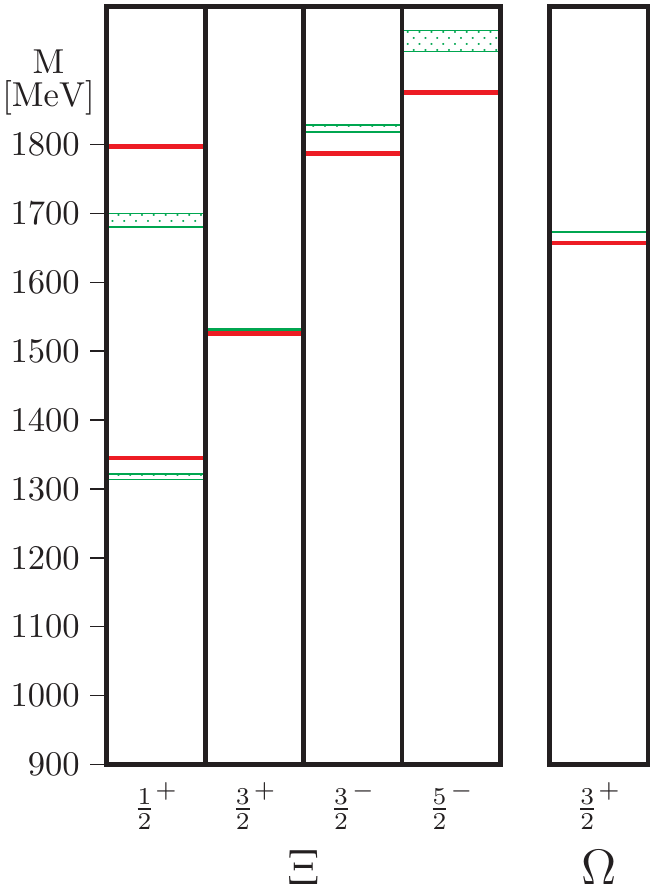}
\caption{\label{fig2} The same as Fig.\ \ref{fig1} for  $\Xi$ and $\Omega$ baryons. }
\end{figure}

The results of our calculations for the strange baryons 
are given in Table \ref{table1}. For comparison, we present the results of the variational calculation 
of Ref.\ \cite{melde08}. The Faddeev results  are slightly lower, mostly by a few MeV. Otherwise the agreement is quite good. 
Figures \ref{fig1} and \ref{fig2} show the agreement with experiment.

 \begin{table}[htb]
 \caption{Here we present our results and compare to the variational calculations of Ref.~\cite{melde08} and to experiments.  All values are given in MeV.}
\label{table1}
 \begin{ruledtabular}
 \begin{tabular}{lcccc}
Baryon & $J^P$ &Faddeev& Ref.\ \cite{melde08}  & Experiment  \\
\hline
$\Lambda(1116)$&$\frac{1}{2}^+$&1133(0)&1136(0)&1116\\
$\Lambda(1405)$&$\frac{1}{2}^-$&1561(0)&1556(0)&1401-1410\\
$\Lambda(1520)$&$\frac{3}{2}^-$&1561(0)&1556(0)&1519-1521\\
$\Lambda(1600)$&$\frac{1}{2}^+$&1607(1)&1625(1)&1560-1700\\
$\Lambda(1670)$&$\frac{1}{2}^-$&1672(1)&1682(1)&1660-1680\\
$\Lambda(1690)$&$\frac{3}{2}^-$&1672(1)&1682(1)&1685-1695\\
$\Lambda(1800)$&$\frac{1}{2}^-$&1777(2)&1778(2)&1720-1850\\
$\Lambda(1810)$&$\frac{1}{2}^+$&1799(2)&1799(2)&1750-1850\\
$\Lambda(1830)$&$\frac{5}{2}^-$&1777(0)&1778(0)&1810-1830\\
\hline
$\Sigma(1193)$&$\frac{1}{2}^+$&1163(0)&1180(0)&1189-1197\\
$\Sigma(1385)$&$\frac{3}{2}^+$&1391(0)&1389(0)&1383-1387\\
$\Sigma[1560]$&$\frac{1}{2}^-$&1666(0)&1677(0)&1546-1576\\
$\Sigma[1620]$&$\frac{1}{2}^-$&1734(1)&1736(1)&1594-1643\\
$\Sigma(1660)$&$\frac{1}{2}^+$&1605(1)&1616(1)&1630-1690\\
$\Sigma(1670)$&$\frac{3}{2}^-$&1666(0)&1677(0)&1665-1685\\
$\Sigma[1690]$&$\frac{3}{2}^+$&1864(1)&1865(1)&1670-1727\\
$\Sigma(1750)$&$\frac{1}{2}^-$&1753(2)&1759(2)&1730-1800\\
$\Sigma(1775)$&$\frac{5}{2}^-$&1734(0)&1736(0)&1770-1780\\
$\Sigma(1880)$&$\frac{1}{2}^+$&1891(2)&1911(2)&1806-2025\\
$\Sigma[1940]$&$\frac{3}{2}^-$&1734(1)&1736(1)&1900-1950\\
$\Sigma$&$\frac{3}{2}^-$&1753(2)&1759(2)&-\\
\hline
$\Xi(1318)$&$\frac{1}{2}^+$&1345(0)&1348(0)&1315-1321\\
$\Xi(1530)$&$\frac{3}{2}^+$&1526(0)&1528(0)&1532-1535\\
$\Xi[1690]$&$\frac{1}{2}^+$&1797(1)&1805(1)&1680-1700\\
$\Xi(1820)$&$\frac{3}{2}^-$&1787(0)&1792(0)&1818-1828\\
$\Xi[1950]$&$\frac{5}{2}^-$&1875(0)&1881(0)&1935-1965\\
\hline
$\Omega$&$\frac{3}{2}^+$&1657(0)&1656(0)&1672.45$\pm$0.29\\
\end{tabular}
\end{ruledtabular}
\end{table}



\section{Summary and conclusions}

In this work we studied the spectrum of strange baryons. We adopted a Goldstone boson exchange model for
the quark-quark interaction and a relativistic form for the kinetic energy. 
We solved the three-body problem by using the Faddeev method. 
The Faddeev method offers several advantages over other more conventional methods that 
solve a Schr\"odinger-type equation. 

In the Faddeev method the Faddeev components possess a simpler structure and they are easier to approximate. 
Also the identity and exchange symmetry of particles are easier to take into account. This is especially important as the symmetry
in the case of strange baryons is delicate. Here we consider the light quarks as identical and treat them in the framework of 
isospin symmetry.  {The light and strange quarks are also identical in the chiral symmetric limit of the theory. However, this symmetry is 
explicitly broken in nature hence the exchange bosons and quarks acquire mass. Additionally the masses of the strange and light quarks
acquire different constituent masses via dynamical and explicit symmetry breaking}. This mixture of exact and broken symmetry can be treated nicely in the two-component version of the Faddeev 
method. In our method the particles $2$ and $3$ are identical and the exchange symmetry is exact. The symmetry between 
particles $1$ and $2$ or $3$ is broken. The fact that it is an exact symmetry in the chiral limit is ensured by the potential form of 
Eq.\ (\ref{octet}), and in particular of Eq.\ (\ref{vus}). This is different in the variational calculation of Ref.\ \cite{melde08}, where the 
wave function was taken to be symmetric with respect to the exchange of all particles irrespective of their mass differences.

\section{Acknowledgements}
We gratefully acknowledge support from the Army High Performance Computing Center consortium ''Hispanic Research and Infrastructure Development Program'' and from the Austrian Science Fund, FWF, through the Doctoral Program on Hadrons in Vacuum, Nuclei, and Stars (FWF DK W1203-N16).

\bibliography{3quark}

\begin{thebibliography}{10}%
\makeatletter
\providecommand \@ifxundefined [1]{%
 \@ifx{#1\undefined}
}%
\providecommand \@ifnum [1]{%
 \ifnum #1\expandafter \@firstoftwo
 \else \expandafter \@secondoftwo
 \fi
}%
\providecommand \@ifx [1]{%
 \ifx #1\expandafter \@firstoftwo
 \else \expandafter \@secondoftwo
 \fi
}%
\providecommand \natexlab [1]{#1}%
\providecommand \enquote  [1]{``#1''}%
\providecommand \bibnamefont  [1]{#1}%
\providecommand \bibfnamefont [1]{#1}%
\providecommand \citenamefont [1]{#1}%
\providecommand \href@noop [0]{\@secondoftwo}%
\providecommand \href [0]{\begingroup \@sanitize@url \@href}%
\providecommand \@href[1]{\@@startlink{#1}\@@href}%
\providecommand \@@href[1]{\endgroup#1\@@endlink}%
\providecommand \@sanitize@url [0]{\catcode `\\12\catcode `\$12\catcode
  `\&12\catcode `\#12\catcode `\^12\catcode `\_12\catcode `\%12\relax}%
\providecommand \@@startlink[1]{}%
\providecommand \@@endlink[0]{}%
\providecommand \url  [0]{\begingroup\@sanitize@url \@url }%
\providecommand \@url [1]{\endgroup\@href {#1}{\urlprefix }}%
\providecommand \urlprefix  [0]{URL }%
\providecommand \Eprint [0]{\href }%
\providecommand \doibase [0]{http://dx.doi.org/}%
\providecommand \selectlanguage [0]{\@gobble}%
\providecommand \bibinfo  [0]{\@secondoftwo}%
\providecommand \bibfield  [0]{\@secondoftwo}%
\providecommand \translation [1]{[#1]}%
\providecommand \BibitemOpen [0]{}%
\providecommand \bibitemStop [0]{}%
\providecommand \bibitemNoStop [0]{.\EOS\space}%
\providecommand \EOS [0]{\spacefactor3000\relax}%
\providecommand \BibitemShut  [1]{\csname bibitem#1\endcsname}%
\let\auto@bib@innerbib\@empty
\bibitem [{\citenamefont {Glozman}\ and\ \citenamefont
  {Riska}(1996)}]{Glozman:1995fu}%
  \BibitemOpen
  \bibfield  {author} {\bibinfo {author} {\bibfnamefont {L.~Y.}\ \bibnamefont
  {Glozman}}\ and\ \bibinfo {author} {\bibfnamefont {D.}~\bibnamefont
  {Riska}},\ }\href {\doibase 10.1016/0370-1573(95)00062-3} {\bibfield
  {journal} {\bibinfo  {journal} {Phys.Rept.}\ }\textbf {\bibinfo {volume}
  {268}},\ \bibinfo {pages} {263} (\bibinfo {year} {1996})},\ \Eprint
  {http://arxiv.org/abs/hep-ph/9505422} {arXiv:hep-ph/9505422 [hep-ph]}
  \BibitemShut {NoStop}%
\bibitem [{\citenamefont {Glozman}\ \emph {et~al.}(1996)\citenamefont
  {Glozman}, \citenamefont {Papp},\ and\ \citenamefont
  {Plessas}}]{Glozman:1996wq}%
  \BibitemOpen
  \bibfield  {author} {\bibinfo {author} {\bibfnamefont {L.~Y.}\ \bibnamefont
  {Glozman}}, \bibinfo {author} {\bibfnamefont {Z.}~\bibnamefont {Papp}}, \
  and\ \bibinfo {author} {\bibfnamefont {W.}~\bibnamefont {Plessas}},\ }\href
  {\doibase 10.1016/0370-2693(96)00610-7} {\bibfield  {journal} {\bibinfo
  {journal} {Phys. Lett.}\ }\textbf {\bibinfo {volume} {B381}},\ \bibinfo
  {pages} {311} (\bibinfo {year} {1996})},\ \Eprint
  {http://arxiv.org/abs/hep-ph/9601353} {arXiv:hep-ph/9601353} \BibitemShut
  {NoStop}%
\bibitem [{\citenamefont {Glozman}\ \emph
  {et~al.}(1998{\natexlab{a}})\citenamefont {Glozman}, \citenamefont {Papp},
  \citenamefont {Plessas}, \citenamefont {Varga},\ and\ \citenamefont
  {Wagenbrunn}}]{Glozman:1997fs}%
  \BibitemOpen
  \bibfield  {author} {\bibinfo {author} {\bibfnamefont {L.~Y.}\ \bibnamefont
  {Glozman}}, \bibinfo {author} {\bibfnamefont {Z.}~\bibnamefont {Papp}},
  \bibinfo {author} {\bibfnamefont {W.}~\bibnamefont {Plessas}}, \bibinfo
  {author} {\bibfnamefont {K.}~\bibnamefont {Varga}}, \ and\ \bibinfo {author}
  {\bibfnamefont {R.~F.}\ \bibnamefont {Wagenbrunn}},\ }\href {\doibase
  10.1103/PhysRevC.57.3406} {\bibfield  {journal} {\bibinfo  {journal} {Phys.
  Rev. C}\ }\textbf {\bibinfo {volume} {57}},\ \bibinfo {pages} {3406}
  (\bibinfo {year} {1998}{\natexlab{a}})}\BibitemShut {NoStop}%
\bibitem [{\citenamefont {Glozman}\ \emph
  {et~al.}(1998{\natexlab{b}})\citenamefont {Glozman}, \citenamefont {Plessas},
  \citenamefont {Varga},\ and\ \citenamefont {Wagenbrunn}}]{Glozman:1997ag}%
  \BibitemOpen
  \bibfield  {author} {\bibinfo {author} {\bibfnamefont {L.~Y.}\ \bibnamefont
  {Glozman}}, \bibinfo {author} {\bibfnamefont {W.}~\bibnamefont {Plessas}},
  \bibinfo {author} {\bibfnamefont {K.}~\bibnamefont {Varga}}, \ and\ \bibinfo
  {author} {\bibfnamefont {R.~F.}\ \bibnamefont {Wagenbrunn}},\ }\href
  {\doibase 10.1103/PhysRevD.58.094030} {\bibfield  {journal} {\bibinfo
  {journal} {Phys. Rev. D}\ }\textbf {\bibinfo {volume} {58}},\ \bibinfo
  {pages} {094030} (\bibinfo {year} {1998}{\natexlab{b}})}\BibitemShut
  {NoStop}%
\bibitem [{\citenamefont {Melde}\ \emph {et~al.}(2008)\citenamefont {Melde},
  \citenamefont {Plessas},\ and\ \citenamefont {Sengl}}]{melde08}%
  \BibitemOpen
  \bibfield  {author} {\bibinfo {author} {\bibfnamefont {T.}~\bibnamefont
  {Melde}}, \bibinfo {author} {\bibfnamefont {W.}~\bibnamefont {Plessas}}, \
  and\ \bibinfo {author} {\bibfnamefont {B.}~\bibnamefont {Sengl}},\ }\href
  {\doibase 10.1103/PhysRevD.77.114002} {\bibfield  {journal} {\bibinfo
  {journal} {Phys. Rev. D}\ }\textbf {\bibinfo {volume} {77}},\ \bibinfo
  {pages} {114002} (\bibinfo {year} {2008})}\BibitemShut {NoStop}%
\bibitem [{\citenamefont {Papp}(1999)}]{Papp:1998yt}%
  \BibitemOpen
  \bibfield  {author} {\bibinfo {author} {\bibfnamefont {Z.}~\bibnamefont
  {Papp}},\ }\href@noop {} {\bibfield  {journal} {\bibinfo  {journal} {Few Body
  Systems}\ }\textbf {\bibinfo {volume} {26}},\ \bibinfo {pages} {99} (\bibinfo
  {year} {1999})}\BibitemShut {NoStop}%
\bibitem [{\citenamefont {Papp}\ \emph {et~al.}(2000)\citenamefont {Papp},
  \citenamefont {Krassnigg},\ and\ \citenamefont {Plessas}}]{Papp:2000kp}%
  \BibitemOpen
  \bibfield  {author} {\bibinfo {author} {\bibfnamefont {Z.}~\bibnamefont
  {Papp}}, \bibinfo {author} {\bibfnamefont {A.}~\bibnamefont {Krassnigg}}, \
  and\ \bibinfo {author} {\bibfnamefont {W.}~\bibnamefont {Plessas}},\ }\href
  {\doibase 10.1103/PhysRevC.62.044004} {\bibfield  {journal} {\bibinfo
  {journal} {Phys. Rev.}\ }\textbf {\bibinfo {volume} {C62}},\ \bibinfo {pages}
  {044004} (\bibinfo {year} {2000})},\ \Eprint
  {http://arxiv.org/abs/nucl-th/0002006} {arXiv:nucl-th/0002006} \BibitemShut
  {NoStop}%
\bibitem [{\citenamefont {McEwen}\ \emph {et~al.}(2010)\citenamefont {McEwen},
  \citenamefont {Day}, \citenamefont {Gonzalez}, \citenamefont {Papp},\ and\
  \citenamefont {Plessas}}]{McEwen:2010sv}%
  \BibitemOpen
  \bibfield  {author} {\bibinfo {author} {\bibfnamefont {J.}~\bibnamefont
  {McEwen}}, \bibinfo {author} {\bibfnamefont {J.}~\bibnamefont {Day}},
  \bibinfo {author} {\bibfnamefont {A.}~\bibnamefont {Gonzalez}}, \bibinfo
  {author} {\bibfnamefont {Z.}~\bibnamefont {Papp}}, \ and\ \bibinfo {author}
  {\bibfnamefont {W.}~\bibnamefont {Plessas}},\ }\href {\doibase
  10.1007/s00601-010-0087-7} {\bibfield  {journal} {\bibinfo  {journal} {Few
  Body Syst.}\ }\textbf {\bibinfo {volume} {47}},\ \bibinfo {pages} {225}
  (\bibinfo {year} {2010})},\ \Eprint {http://arxiv.org/abs/1001.4062}
  {arXiv:1001.4062 [nucl-th]} \BibitemShut {NoStop}%
\bibitem [{\citenamefont {Schellingerhout}()}]{Schellingerhout:1995gn}%
  \BibitemOpen
  \bibfield  {author} {\bibinfo {author} {\bibfnamefont {N.~W.}\ \bibnamefont
  {Schellingerhout}},\ }\href@noop {} {\ }\bibinfo {note} {RX-1544
  (GRONINGEN)}\BibitemShut {NoStop}%
\bibitem [{\citenamefont {Nakamura}\ \emph {et~al.}(2010)\citenamefont
  {Nakamura} \emph {et~al.}}]{pdg}%
  \BibitemOpen
  \bibfield  {author} {\bibinfo {author} {\bibfnamefont {K.}~\bibnamefont
  {Nakamura}} \emph {et~al.} (\bibinfo {collaboration} {Particle Data Group}),\
  }\href {\doibase 10.1088/0954-3899/37/7A/075021} {\bibfield  {journal}
  {\bibinfo  {journal} {J.Phys.G}\ }\textbf {\bibinfo {volume} {G37}},\
  \bibinfo {pages} {075021} (\bibinfo {year} {2010})}\BibitemShut {NoStop}%
\end{thebibliography}%

\end{document}